\documentclass[twoside,twocolumn,9pt]{article}
\usepackage{amsmath}
\usepackage{amssymb}
\usepackage{extsizes}
\usepackage[version=3]{mhchem}
\usepackage[left=1.5cm, right=1.5cm, top=1.785cm, bottom=2.0cm]{geometry}
\usepackage{balance}
\usepackage{times,mathptmx}
\usepackage{sectsty}
\usepackage{graphicx} 
\usepackage{lastpage}
\usepackage[format=plain,justification=raggedright,singlelinecheck=false,font={stretch=1.125,small,sf},labelfont=bf,labelsep=space]{caption}
\usepackage{float}
\usepackage{fancyhdr}
\usepackage{fnpos}
\usepackage[english]{babel}
\usepackage{array}
\usepackage{droidsans}
\usepackage{charter}
\usepackage[T1]{fontenc}
\usepackage[usenames,dvipsnames]{xcolor}
\usepackage{setspace}
\usepackage[compact]{titlesec}
\usepackage[scientific-notation=true]{siunitx}
\usepackage[super,sort&compress,comma]{natbib} 
\usepackage{booktabs}
\usepackage{multirow} \usepackage{multicol}
\usepackage{url}
\usepackage{appendix}
\definecolor{cream}{RGB}{222,217,201}

\pdfoutput=1
\usepackage{hyperref}
\hypersetup{
    colorlinks=true,
    linkcolor=black,
    citecolor=black,
    filecolor=black,
    urlcolor=black,
}

\begin{document}



\makeFNbottom
\makeatletter
\renewcommand\LARGE{\@setfontsize\LARGE{15pt}{17}}
\renewcommand\Large{\@setfontsize\Large{12pt}{14}}
\renewcommand\large{\@setfontsize\large{10pt}{12}}
\renewcommand\footnotesize{\@setfontsize\footnotesize{7pt}{10}}
\makeatother

\renewcommand{\thefootnote}{\fnsymbol{footnote}}
\renewcommand\footnoterule{\vspace*{1pt}%
\color{cream}\hrule width 3.5in height 0.4pt \color{black}\vspace*{5pt}} 
\setcounter{secnumdepth}{5}

\makeatletter 
\renewcommand\@biblabel[1]{#1}            
\renewcommand\@makefntext[1]%
{\noindent\makebox[0pt][r]{\@thefnmark\,}#1}
\makeatother 
\renewcommand{\figurename}{\small{Fig.}~}
\sectionfont{\sffamily\Large}
\subsectionfont{\normalsize}
\subsubsectionfont{\bf}
\setstretch{1.125} 
\setlength{\skip\footins}{0.8cm}
\setlength{\footnotesep}{0.25cm}
\setlength{\jot}{10pt}
\titlespacing*{\section}{0pt}{4pt}{4pt}
\titlespacing*{\subsection}{0pt}{15pt}{1pt}

\setlength{\arrayrulewidth}{1pt}
\setlength{\columnsep}{6.5mm}
\setlength\bibsep{1pt}

\makeatletter 
\newlength{\figrulesep} 
\setlength{\figrulesep}{0.5\textfloatsep} 

\newcommand{\topfigrule}{\vspace*{-1pt}%
\noindent{\color{cream}\rule[-\figrulesep]{\columnwidth}{1.5pt}} }

\newcommand{\botfigrule}{\vspace*{-2pt}%
\noindent{\color{cream}\rule[\figrulesep]{\columnwidth}{1.5pt}} }

\newcommand{\dblfigrule}{\vspace*{-1pt}%
\noindent{\color{cream}\rule[-\figrulesep]{\textwidth}{1.5pt}} }

\makeatother

\twocolumn[
  \begin{@twocolumnfalse}
\sffamily

\LARGE{\textbf{A universal chemical potential for sulfur vapours\(^\dag\)}} \\
\vspace{0.3cm} \\
\noindent\large{
Adam J. Jackson\textit{\(^{a}\)}, Davide Tiana\textit{\(^{a,\ddag}\)} and Aron Walsh\(^{\ast}\)\textit{\(^{a,b}\)}} \\

\normalsize{
The unusual chemistry of sulfur is illustrated by the tendency for catenation. 
Sulfur forms a range of open and closed S\(_{n}\) species in the gas phase, which has led to speculation on the composition of sulfur vapours as a function of temperature and pressure for over a century. 
Unlike elemental gases such as O\(_{\text{2}}\) and N\(_{\text{2}}\), there is no widely accepted thermodynamic potential for sulfur. 
Here we combine a first-principles global structure search for the low energy clusters from S\(_{\text{2}}\) to S\(_{\text{8}}\) with a thermodynamic model for the mixed-allotrope system,
including the Gibbs free energy for all gas-phase sulfur on an atomic basis. 
A strongly pressure-dependent transition from a mixture dominant in S\(_{\text{2}}\) to S\(_{\text{8}}\) is identified. 
A universal chemical potential function, \(\mu_{\mathrm{S}}(T,P)\), is proposed with wide utility in modelling sulfurisation processes including the formation and annealing of metal chalcogenide
semiconductors.
} \\


 \end{@twocolumnfalse} \vspace{0.6cm}

  ]

\renewcommand*\rmdefault{bch}\normalfont\upshape
\rmfamily
\section*{}
\vspace{-1cm}


\footnotetext{$^{a}$\emph{Centre for Sustainable Chemical Technologies and Dept. of Chemistry, University of Bath, Claverton Down, Bath BA2 7AY, UK}}

\footnotetext{$^{b}$\emph{Global E$^3$ Institute and Department of Materials Science and Engineering, Yonsei University, Seoul 120-749, Korea}}

\footnotetext{$^{\ddag}$ Current: \emph{EPFL Valais Wallis, EPFL LSMO, Rue de l'Industrie 17, Case postale 440, CH-1951 Sion, Switzerland}}

\footnotetext{\dag~Electronic Supplementary Information (ESI) available: Tabulated free energy and enthalpy data. See DOI: 10.1039/C5SC03088A. Additional data and code available in external repositories with DOIs: 10.5281/zenodo.28536; 10.6084/m9.figshare.151373; 10.6084/m9.figshare.1513833. See Data Access Statement for more information.}

\section{Introduction}
\label{sec:orgheadline1}

Sulfur is an abundant resource exploited by industry on a scale of tens of millions of tonnes per year.\cite{Nehb2000}
While it may be found in its elemental form, the primary industrial source is hydrogen sulfide, a byproduct of the oil and gas industry.
The vast majority of industrial sulfur is converted to sulfuric acid or sulfur dioxide before further use;
this may explain the surprising shortage of data in the thermochemical literature regarding the vapour phase of elemental sulfur.

Historically, the thermochemistry of sulfur has been studied experimentally and has been understood to be associated with a variable composition for over a century;
Lewis and Randall remarked in 1914 that "no other element is known to occur in as many different forms as sulfur" while studying the free energy of a number of these forms.\cite{Lewis1914}
(Carbon now has a higher number of known allotropes but the majority of these are not naturally-occuring.)
However, contemporary reference data for sulfur still does not present a complete picture;
the NIST-JANAF Thermochemical Tables (1998) give thermochemical data for two solid phases, one liquid phase, the ions S\(^{\text{+}}\) and S\(^{\text{-}}\) and eight gas allotropes S\(_{\text{1-8}}\).\cite{Chase1998} Of these, only S\(_{\text{2}}\) and S\(_{\text{8}}\) are from spectroscopic data.
The allotropes S\(_{\text{3-7}}\) are assumed to exist and are assigned energies following an interpolation scheme suggested by Rau \emph{et al.} (1966), which also makes use of experimental data for S\(_{\text{6}}\).\cite{Rau1973a}
That paper rules out the significant presence of tautomers, finding little evidence of a tautomer contribution and assuming that they have relatively high energy.
The authors generally reserve speculation on the actual structures of the components of their equilibrium model.

In recent years considerable attention has turned to metal chalcogenides;
II-VI semiconductors such as ZnS, CdS, PbS are widely studied in many contexts.\cite{Yu2003}
Copper indium gallium selenides (CIGS) and cadmium telluride (CdTe) are used as the basis for "second-generation" thin-film photovoltaic devices, and have seen a dramatic rise in production.
Cu\(_{\text{2}}\)ZnSn(S,Se)\(_{\text{4}}\) (CZTS) and Cu\(_{\text{2}}\)SnS\(_{\text{3}}\) (CTS) devices have so far struggled to match these materials in terms of energy conversion efficiencies, but hold significant long-term promise due to their use of highly abundant elements; such availability is a prerequisite for terawatt-scale photovoltaics.\cite{Berg2012}
As such, thin-film processing in sulfur atmospheres is of considerable interest, as the inherent safety of industrial processing may be improved by eliminating the use of toxic H\(_{\text{2}}\)S.
In addition to chalcogen annealing, which is used to increase grain size, substitute other elements or directly form chalcogenides from elements, high-quality single-crystal samples may be produced using chemical vapour transport of elemental chalogens.\cite{Lichtenstriger1961, Colombara2013, Burton2013}
Previous work on the thermodynamics of such processing has tended to assume that sulfur adopts one particular gaseous allotrope (either S\(_{\text{2}}\) or S\(_{\text{8}}\)), 
but the validity of this assumption has not been explored in depth.\cite{Jackson2014,Kosyak2013,Scragg2011a}
It is undermined however by the model derived by Rau \emph{et al.}, which predicts that no one component makes more than 50\% of the gas mixture at temperatures between 800-1100K.\cite{Rau1973a}

Mass spectrometry at a relatively mild 105\(^{\circ}\)C has observed a series of charged clusters with the form (S\(_{\text{8n}}\))\(^{\text{+}}\).\cite{Martin1984}
In the mid 1980s, a number of cyclic allotropes had been identified by crystallisation and X-ray diffraction, but this only covered the range \(n\) = 6 -- 20.\cite{Steudel1984} 
An \emph{ab initio} study was carried out for S\(_{\text{2}}\) through to S\(_{\text{13}}\) in an early application of the Car-Parrinello simulated annealing method.\cite{Hohl1988}
Energies were calculated using density-functional theory with the local density approximation (LDA).
While limited by the inherent difficulties in exploring the entire potential energy surface of the atomic positions, 
this thorough study generated 21 allotropes, finding a local maximum in the atomisation energy at \(n=8\).
A later (1990) paper used coupled-cluster electronic structure calculations to study the proposed tautomers of S\(_{\text{4}}\) in depth, concluding that the planar structure with \(C_{2v}\) symmetry is lowest in energy, with a trans (\(C_{2h}\)) structure also visible in experimental spectra;
a more recent \emph{ab initio} study reached similar conclusions regarding stability while challenging the spectroscopic assignment of the phases.\cite{Quelch1990, Wong2003}
The \(C_{2v}\) structure was ruled out in the simulated annealing study with LDA, although the authors noted the experimental evidence for its existence.\cite{Hohl1988}
A 2003 review by \citet{Steudel2003} collects more recent data, including both experimental and theoretical studies of vapour-phase allotropes; this review notes the weakness of the widespread assumption that each size is represented by a single species.\cite{Steudel2003}
The work compares several sets of enthalpies relative to S\(_{\text{8}}\) that have been obtained experimentally;
variability is high for the smaller allotropes while there is fairly good agreement for the larger allotropes.
Studies are generally carried out at a single temperature, such that the temperature and pressure dependence of the thermochemistry must be derived from statistical mechanics and analysis of vibrational information.

In this study, we develop a set of structures for S\(_{\text{2}}\)-S\(_{\text{8}}\), compute their Gibbs free energy from first-principles and with empirical corrections, and solve the temperature-dependent chemical potential to describe the gaseous mixture.
The potential function will be important for quantitative investigations of defect formation and phase stability in metal sulfide materials.

\section{Methods}
\label{sec:orgheadline9}

\subsection{Density functional theory}
\label{sec:orgheadline2}
Energies and forces of arbitrary clusters of sulfur atoms were computed within Kohn-Sham density-functional theory (DFT).\cite{Kohn1965}
A range of exchange-correlation functionals were used in this work: PBE is a popular and elegant implentation of the Generalised Gradient Approximation (GGA) and PBEsol restores a periodic exchange contribution leading to improved performance for solids;\cite{Perdew1996,Perdew2008} 
B3LYP\footnote{Note that the implementation of B3LYP in FHI-aims uses a parameterisation of the local density contribution based on the Random Phase Approximation in order to match values obtained with Gaussian, another quantum chemistry code.\cite{Hertwig1997}} is a widely-used "hybrid" functional which combines pre-existing gradient corrections with "exact" Hartree-Fock exchange;\cite{Becke1993} PBE0 is applies similar principles to the parameter-free PBE functional.\cite{Adamo1999}
(While PBE is generally preferred to PBEsol for molecular calculations, PBEsol was included in this study for its compatibility with other all-electron work using this functional.)

Calculations for the evolutionary algorithm search used the Vienna Ab Initio Simulations Package (VASP)  with the PBE exchange-correlation functional and a plane-wave basis set with a 500 eV energy cutoff.\cite{Kresse1996b,Kresse1996c} 
As calculations in VASP employ a periodic boundary condition, orthorhombic bounding boxes were employed with 10 \(\AA\) of vacuum between each molecule and its periodic images.
Electronic structure iteration used only the \(\Gamma\)-point of this large cell.

Further calculations used the Fritz Haber Institute ab initio molecular simulations package (FHI-aims) to carry out all-electron DFT calculations with numerically-tabulated basis sets.\cite{Blum2009,Havu2009}
All calculations were open-shell with S\(_2\) adopting its low-energy triplet spin configuration.
The recommended "tight" basis set was employed for initial relaxation and study with PBEsol, which extends the minimal set of occupied orbitals with 6 additional functions.
This was extended further to the full "tier 2" set of 9 additional functions for calculations with the LDA, PBE0, and B3LYP functionals.

\subsection{Global structure search}
\label{sec:orgheadline3}
Global structure optimisation was carried out with the USPEX package, which was originally developed for crystalline systems and has been adapted for use with clusters.\cite{Oganov2006,Oganov2011,Lyakhov2013}
At this stage, molecules with \(n>8\) were disregarded, as experimental results anticipate high- and low-temperature limits dominated by S\(_{\text{2}}\) and S\(_{\text{8}}\), respectively.
Clusters were generated for S\(_{\text{3-7}}\), and refined with an evolutionary algorithm to minimise the ground-state energy until a number of seemingly distinct clusters were identified by inspection.
The atomic positions of these clusters were then optimised in FHI-aims calculations with PBEsol, using the BFGS algorithm to minimise the atomic forces to less than 10\(^{\text{-4}}\)~eV~\AA{}\(^{\text{-1}}\)  and converge energy to within 10\(^{\text{-6}}\)~eV.
Point groups were assigned to the structures using Materials Studio version 6.0, a proprietary package developed by Accelrys.

\subsection{Vibrational frequencies}
\label{sec:orgheadline4}
Vibrational frequencies were calculated within the harmonic approximation by making finite displacements to each atomic position to obtain the local potential wells, and diagonalising the resulting dynamical matrix to obtain the normal modes and their frequencies.
This is implemented as a script and diagonalisation routine provided with FHI-aims.

Improved vibrational frequencies may be obtained by applying an empirically-derived scale factor to the vibrational eigenvalues computed using DFT;
collections of such scale factors have been published for large test-sets of molecules.\cite{Merrick2007,Alecu2010}
The use of these factors is somewhat problematic when creating a systematic, transferable set of data but offers an opportunity to create the most realistic thermochemical model possible.
Given that the calculations in this work involve a more limited subset of atomic interactions, we choose to fit a scaling factor to the experimentally-reported frequencies of S\(_8\) and S\(_2\).

\subsection{Thermochemistry}
\label{sec:orgheadline8}
\subsubsection{Thermochemistry of individual gas species.~~}
\label{sec:orgheadline5}
\label{SEC:species-thermochem}
Thermochemical properties were calculated within the ideal gas, rigid-rotor and harmonic vibration approximations.
A set of textbook equations forms the chemical potential \(\mu\) for a nonlinear molecule from the ground-state electronic energy \(E_0\) given a set of vibrational energies \(\mathbf{\epsilon}\), the rotational constant \(\sigma\), moment of inertia \(I\)
\begin{align}
\mu &= E_0 + E_\text{ZPE} + \int^T_0 C_{v} + k_B T - T S \\
\intertext{where}
C_v &= C_{v,\textrm{trans}} + C_{v,\textrm{vib}} + C_{v,\textrm{rot}}\\
\int^T_0 C_v &\approx \frac{3}{2} k_B + \sum_i \frac{\epsilon_i}{\exp(\epsilon_i / k_B T) - 1} + \frac{3}{2} k_B \\
S &= S_\textrm{vib} + S_\textrm{trans} + S_\textrm{rot} \\
&= \sum_i \left[ \frac{\epsilon_i / k_B T}{\exp (\epsilon_i / k_B T) - 1} - \ln \left( 1-\exp (-\epsilon_i / k_B T)\right) \right] \nonumber \\
&\quad + k_B  \left[ \ln \left( \frac{2 \pi m k_B T}{h^2} \right)^{\tfrac{3}{2}} \frac{k_B T}{P_{\textrm{ref}}}p + \frac{5}{2} \right] \nonumber \\
&\quad + k_B \left[ \ln \frac{\sqrt{\pi \prod_i I_i}}{\sigma} \left( \frac{8 \pi^2 k_B T}{h^2} \right)^{\tfrac{3}{2}} + \frac{3}{2}  \right]. \nonumber \\
\end{align}
These were applied as implemented in the Atomic Simulation Environment (ASE) Python package.\cite{Bahn2002}
(Note that the expressions for monatomic and linear molecules are slightly different.)
The rotational constants \(\sigma\) were assigned from the point groups.

\subsubsection{Reference energies.~~}
\label{sec:orgheadline6}
A number of \emph{ab initio} methods have been applied. In order to compare the energies, a reference point is needed.
Conventionally the enthalpy of the ground state is zero; however, in this case the ground state phase \(\alpha\)-sulfur is relatively expensive to compute.
We therefore use the experimental sublimation enthalpy \(\Delta H_{sub} = \tfrac{1}{8} H_{\textrm{S}_8} - H_{\textrm{S}_{\alpha}}\) to obtain a reference from the calculated enthalpy of S\(_8\):
\begin{align}
\Delta H_{\text{S}_x} &= H_{\text{S}_x} - x H_{\text{S}_\alpha} \\
\Delta H_{\text{S}_x} &= H_{\text{S}_x} - x \left( \frac{H_{\text{S}_8}}{8} + H_{\text{S}_\alpha} - \frac{H_{\text{S}_8}}{8} \right) \\
\Delta H_{\text{S}_x} &= H_{\text{S}_x} - x \left( \frac{H_{\text{S}_8}}{8} - \Delta H_{sub} \right)
\end{align}
The preferred experimental value for \(\Delta H_{sub}\) is \(100.416 / 8 =  12.552\) kJ mol\(^{-1}\), from experiments at 298K.\cite{Chase1998}
Note that the physical system does not in fact sublime at high temperatures, but passes through a molten phase. Nonetheless, it is more practical (and perfectly valid) to retain \(\alpha\)-S as the reference state over the whole temperature range studied.

\subsubsection{Equilibrium modelling.~~}
\label{sec:orgheadline7}
\label{SEC:eqm_derivation}
The following derivation closely follows the approach and notation of Ref.~\citenum{Smith1982},
which describes a generalised "non-stoichiometric method" for solving chemical equilibria.
This approach is well-established and based on key work in Refs.~\citenum{Brinkley1946,Brinkley1947,White1967}.

We attempt to minimise the Gibbs free energy
\begin{align}
\min G(\mathbf{n}) &= \sum^N_{i=1} n_i \mu_i \label{eq:minG}\\
\intertext{subject to the mass balance constraint}
\sum^N_{i=1} a_{i} n_i &= b \label{eq:mass_balance}
\end{align}
where \(N\) is the number of unique species \(i\) with stoichiometric coefficient \(a_i\); \(n\) is the quantity of species \(i\) and \(b\) is the total number of sulfur atoms.
The classic approach for a constrained optimisation is the method of Lagrange multipliers.
The Lagrangian is formed
\begin{align}
{\cal L}(\mathbf{n},\lambda) &= \sum^N_{i=1} n_i \mu_i + \lambda \left( b - \sum^N_{i=1} a_i n_i \right)\\
\intertext{and differentiated to form a set of equations definining the equilibrium state.}
\left(\frac{\partial \cal{L}}{\partial n_i}\right)_{n_{j\ne i, \lambda}} &=
\mu_i - a_i \lambda &= 0 \label{eq:eqm-cond1}\\
\intertext{and}
\left(\frac{\partial \cal{L}}{\partial \lambda} \right)_\mathbf{n} &=
b - \sum^N_{i=1} a_i n_i &= 0. \label{eq:eqm-cond2}
\end{align}
The species chemical potential \(\mu_i\) calculated as in Section~\ref{SEC:species-thermochem} is a function of both temperature and the partial pressure \(p_i = P \frac{n_i}{n_t}\) where \(P\) is the total pressure and the total quantity \(n_t = \sum^N_i n_i\).
The temperature dependence is complex and we are willing to solve the equilibrium at each temperature of interest, so we form a temperature-dependent standard free energy at a reference pressure \(P^\circ\), \(\mu_i^\circ(T) = \mu_i(T,P^\circ)\).
\begin{align}
    \mu_i(T,P,\mathbf{n}) &= \mu_i^\circ (T) + R T \ln \left(\frac{p_i}{P^\circ} \right) \\
    &= \mu_i^\circ (T) + R T \ln\left(\frac{n_i}{n_t} \frac{P}{P^\circ} \right) \\
    &= \mu_i^\circ (T) + R T \ln \left( \frac{P}{P^\circ} \right) + R T \ln\left(\frac{n_i}{n_t} \right) \label{eq:ideal_mu}
\end{align}
From here we drop the parenthetical indication that \(\mu_i^\circ\) is a function of temperature,
and define the unit of pressure as the reference pressure, such that \(P^\circ = 1\).
Substituting (\ref{eq:ideal_mu}) into (\ref{eq:eqm-cond1}), we obtain
\begin{align}
    \mu_i^\circ + R T \ln\left(\frac{n_i}{n_t} P \right) - a_i\lambda &= 0 \\
    \ln\left( \frac{n_i}{n_t}P \right) &= \frac{a_i\lambda - \mu_i^\circ}{R T} \label{eq:1}\\
\intertext{and summing over $i$}
P &= \sum^N_{i=1} \exp\left(\frac{a_i\lambda - \mu_i^\circ}{RT}\right).
\end{align}
The only unknown variable in this expression is \(\lambda\); 
rearranging slightly we form a polynomial which is suitable for solving by standard numerical methods.
The method employed in this work is the Levenberg-Marquardt least-squares algorithm, as implemented in Scipy.\cite{Marquardt1963,Scipy2001} 
\begin{equation}
\sum^N_{i=1} \exp \left( \frac{-\mu_i^\circ}{RT} \right) \left[ \exp \left( \frac{\lambda}{RT} \right)\right]^{a_i} -P = 0 \label{eq:lagrange_polynomial} \\
\end{equation}
\noindent To recover the composition \(\mathbf{n}\), we rearrange (\ref{eq:1}):
\begin{align}
n_i &= \frac{n_t}{P} \exp\left( \frac{a_i\lambda}{RT} \right) \exp\left(\frac{-\mu_i^\circ}{R T}\right) \label{eq:3} \\
\intertext{and substitute into the second equilibrium condition (\ref{eq:mass_balance}) to obtain}
b &= \frac{n_t}{P}\sum\limits^N_{i=1} a_i \exp \left( \frac{a_i \lambda -\mu_i^\circ}{RT}\right) \label{eq:2}\\
\intertext{combining (\ref{eq:3}) and (\ref{eq:2}) we eliminate $n_t$}
\frac{n_i}{b} &= \frac{\exp\left(\frac{a_i \lambda - \mu_0}{RT} \right)}{\sum\limits^N_{i=1} a_i \exp \left(\frac{a_i \lambda - \mu_0}{RT} \right)} \\
\intertext{
and clean up the notation by denoting $\exp \left( \frac{a_i \lambda - \mu_0}{RT}\right)$ as $\Phi_i$}
\frac{n_i}{b} &= \frac{\Phi_i}{\sum\limits^N_{i=1} a_i \Phi_i}. \label{eq:phi_frac}
\end{align}
Finally, to obtain the chemical potential of the mixture we note from (\ref{eq:eqm-cond1}) that 
\(\frac{\mu_i}{a_i}=\lambda\) for all \(i\).
Therefore
\begin{equation}
\lambda = \mu_{\text{S}},
\end{equation}
the normalised chemical potential of sulfur vapour on an atom basis.
(A mathematical derivation is given in Appendix~\ref{SEC:lagrange_gibbs_proof}.)

\section{Results}
\label{sec:orgheadline23}

\subsection{Sulfur allotropes}
\label{sec:orgheadline18}
   \label{sec:results.types}
A variety of candidate structures were generated in the evolutionary algorithm study with the PBE functional.
The low-energy candidates following geometry optimisation are discussed in this section.

\begin{figure*}
\centering
\includegraphics[width=\textwidth]{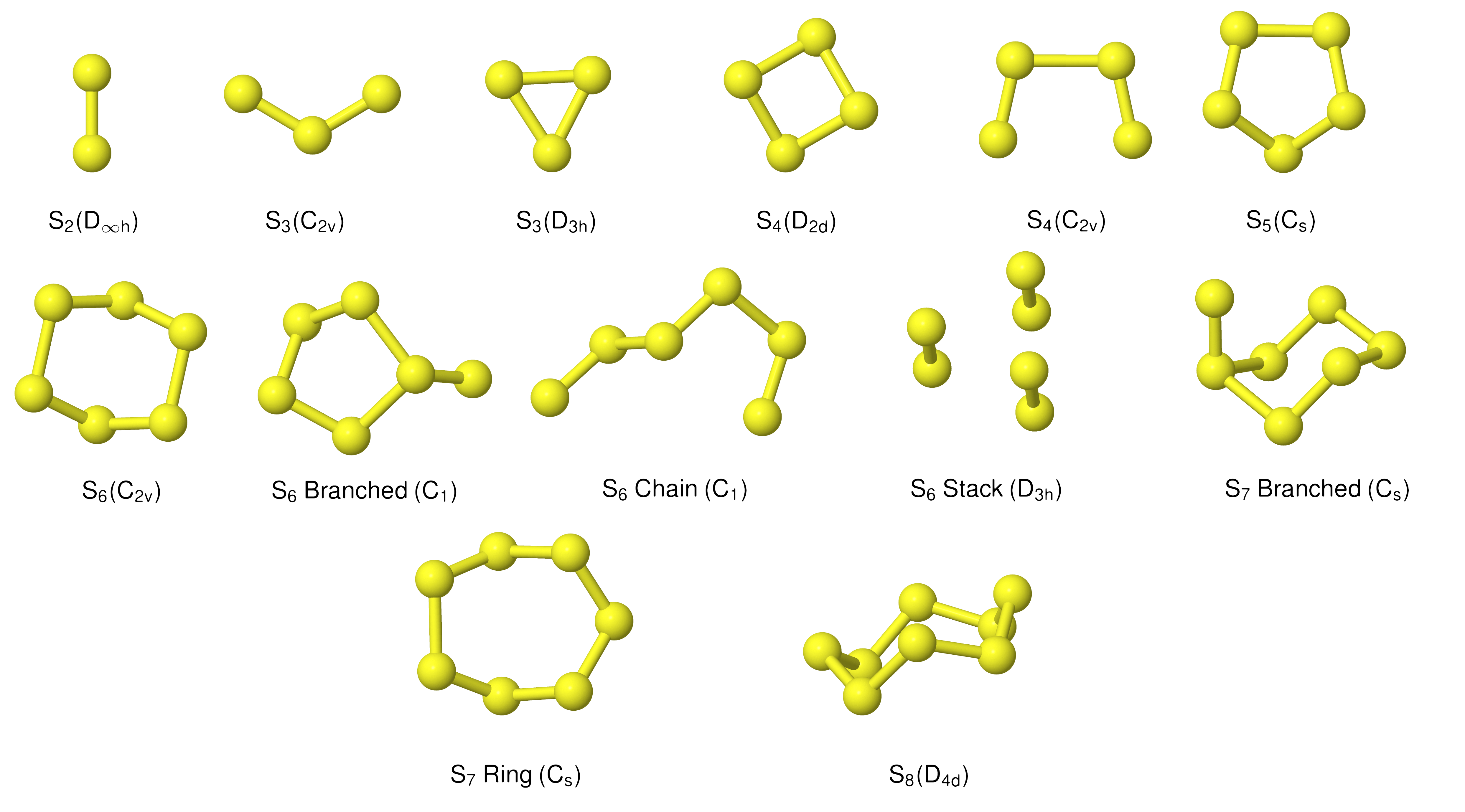}
\caption{\label{fig:S-montage}
Predicted low-energy sulfur clusters with symmetry assignment}
\end{figure*}

\subsubsection{S\(_{\text{2}}\).~~}
\label{sec:orgheadline10}
Diatomic sulfur has the point group \(D_{\infty h}\), in common with other homonuclear diatomics.
The atoms were initially set 2 \AA{} apart, and relaxed to a bond length of 1.91 \AA{}.
Studies with other functionals were relaxed either from this distance or from 2 \AA{}.
The resulting bond lengths are given in Table~\ref{tbl:S2_r}.

\begin{table}[htb]
\small
\caption{\label{tbl:S2_r}
Calculated and experimental bond length \(r\) in S\(_{\text{2}}\). Experimental value is NIST/JANAF-recommended distance.\cite{Chase1998}}
\centering
\begin{tabular}{lr}
\toprule
DFT functional & \(r\) / \AA{}\\
\midrule
PBE & 1.911\\
PBEsol & 1.903\\
LDA & 1.895\\
PBE0 & 1.884\\
\midrule
Experiment & 1.889\\
\bottomrule
\end{tabular}
\end{table}

\subsubsection{S\(_{\text{3}}\).~~}
\label{sec:orgheadline11}

The evolutionary algorithm process eliminated all but a \(C_{2v}\) non-linear chain for S\(_3\).
This corresponds to "thiozone", which has a well-characterised structure by rotational spectroscopy (bond length 1.917(1)~\AA{} and angle 117.36(6)\(^\circ\); the values from optimisation with PBE0 in this study are 1.901~\AA{} and 118.2$^{\circ}$).\cite{McCarthy2004}
We have also considered the simple triangular allotrope, which is \(\sim 0.5\) eV higher in ground-state energy.

\subsubsection{S\(_{\text{4}}\).~~}
\label{sec:orgheadline12}
A range of branched and cyclic structures were generated in the evolutionary algorithm.
The structures included in the equilibrium modelling are shown in Fig.~\ref{fig:S-montage}.
The lowest-energy structure identified was the `eclipsed' C\(_{\text{2v}}\) chain; 
this is in agreement with the high-level theoretical studies in Ref.~\citenum{Quelch1990,Wong2003}.
These studies identified a `trans' \(C_{2h}\) structure as being likely to exist; 
there is some spectroscopic evidence for the viability of this isomer as well as a branched chain,
but we were not able to reproduce stable structures corresponding to these allotropes through geometry optimisation.\cite{Boumedien1999,Hassanzadeh1992}
Various cyclic and tetrahedral candidate structures yielded a relatively flat puckered ring with \(D_{2d}\) symmetry.

\subsubsection{S\(_{\text{5}}\).~~}
\label{sec:orgheadline13}
Although a wide range of branched and chain structures were generated, the main candidate is the 5-membered ring with \(C_{s}\) symmetry.

\subsubsection{S\(_{\text{6}}\).~~}
\label{sec:orgheadline14}
In addition to a cyclic \(C_{2v}\) allotrope, relatively low-energy branched and chain variations were identified.
Of considerable interest is also a structure which may be viewed as a stack of two S\(_3\) cycles, or alternatively as a cluster of S\(_2\) diatoms.
This appears to be the \(D_{3h}\) "prism" structure identified by by \citet{Wong2004}; the characteristic S-S bond lengths from that study were 190.1 and 276.2 pm, while the corresponding average distances from optimisation with the same hybrid XC functional (B3LYP) in this work were 189.0 and 275.7 pm. 
It is worth stressing that no explicit dispersion terms were included in any of the electronic structure calculations.

\subsubsection{S\(_{\text{7}}\).~~}
\label{sec:orgheadline15}
The evolutionary algorithm results rapidly provided the same \(C_s\) cyclic structure as that obtained by energy minimisation from a regular polygon.
A branched structure, generated early in the progress of the algorithm, was also selected as an interesting alternative to include. 
This was about 1 eV lower in energy than the other candidates at that stage.
Geometry optimisation by force relaxation yielded a compact structure, also with \(C_s\) (mirror-plane) symmetry.

\subsubsection{S\(_{\text{8}}\).~~}
\label{sec:orgheadline16}
No evolutionary algorithm study was applied for S\(_8\), as its ring structure is quite well-known.
The initial geometry was extracted from the crystal structure for the condensed \(\alpha\)-S phase used in a previous study,\cite{Burton2012a}
and relaxed to form an isolated \(D_{4d}\) ring.

\subsubsection{Ground-state energies.~~}
\label{sec:orgheadline17}
\begin{figure}[htb]
\centering
\includegraphics[width=.9\linewidth]{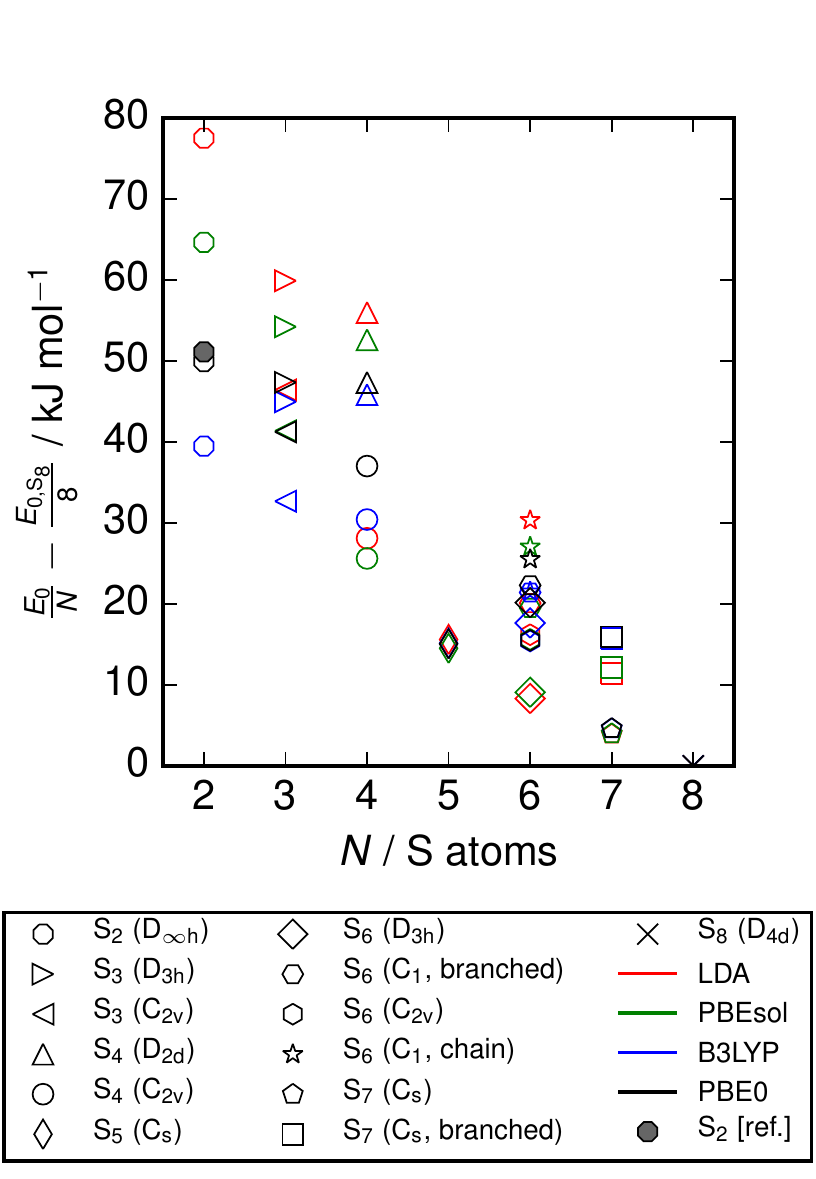}
\caption{\label{fig:energies}
Ground-state energies from DFT of clusters included in study. Energies are relative to the energy for S\(_8\) with each functional, and normalised to the number of atoms. A point is also included from reference data\cite{Chase1998}; this is derived from the enthalpies of formation at zero temperature, based on spectroscopic observations and equilibrium studies. While the energies from different exchange-correlation functionals diverge across the series, the S\(_2\) energy from PBE0 calculations agrees closely with this reference data.}
\end{figure}

An inspection of the ground-state energies from DFT reveals a trend of smoothly decreasing energy per atom with cluster size for the minimum-energy configuration at each size (Fig.~\ref{fig:energies}).
The variation within the clusters included at each size is of the order 10 kJ mol\(^{-1}\) atom\(^{-1}\), which is comparable to the energy difference between neigbouring cluster sizes.

\subsection{Vibrational properties}
\label{sec:orgheadline20}
Vibrational frequencies were calculated for all of the allotropes listed in section \ref{sec:results.types}; frequencies for S\(_2\) and S\(_8\) are listed in Table~\ref{tbl:frequencies}.

\begin{figure}[htb]
\centering
\includegraphics[width=0.4\textwidth]{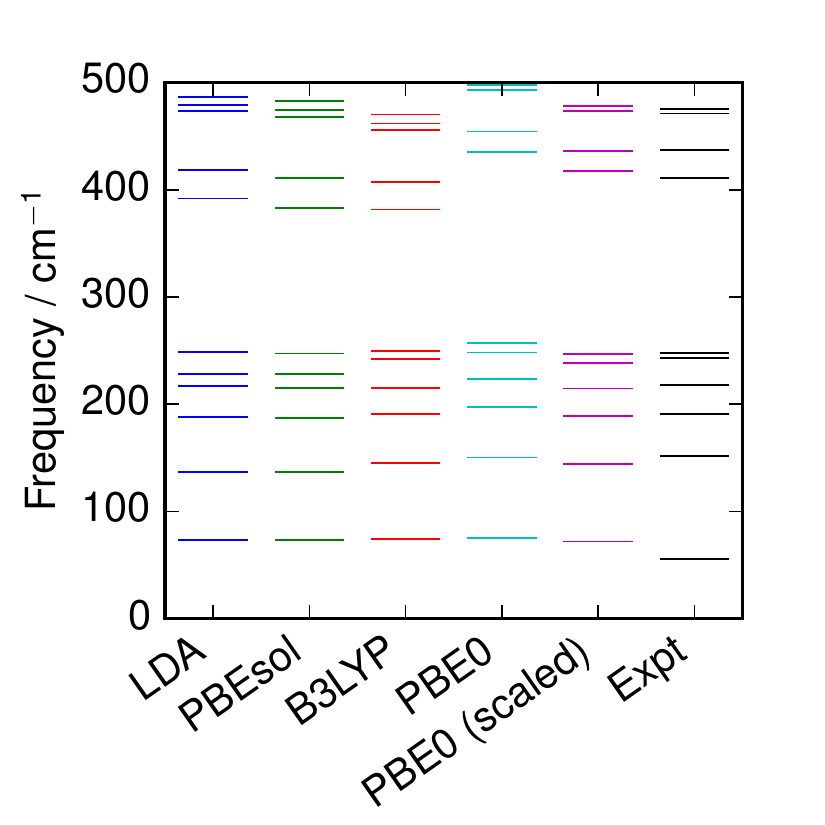}
\caption{\label{fig:empirical_freqs}
Vibrational frequencies of S\(_8\) calculated with various DFT functionals, compared with recommended experimental values.\cite{Chase1998}}
\end{figure}

\begin{table}[htb]
\small
\caption{\label{tbl:frequencies}
Calculated and experimental vibrational frequencies for S\(_{\text{2}}\) and S\(_{\text{8}}\).\cite{Chase1998} All frequencies in cm\(^{\text{-1}}\).}
\centering
\begin{tabular}{lrrrrrr}
\toprule
 & LDA & PBEsol & PBE0 & PBE0 & B3LYP & Expt\\
 &  &  &  & (scaled) &  & \\
\midrule
S\(_{\text{2}}\) & 716 & 713 & 751 & 721 & 714 & 724\\
\midrule
S\(_{\text{8}}\) & 73 & 73 & 74 & 71 & 74 & 56\\
 & 73 & 73 & 75 & 72 & 74 & 56\\
 & 136 & 136 & 150 & 144 & 145 & 152\\
 & 136 & 136 & 150 & 144 & 145 & 152\\
 & 188 & 187 & 197 & 189 & 191 & 191\\
 & 188 & 187 & 197 & 189 & 191 & 191\\
 & 217 & 215 & 223 & 214 & 214 & 218\\
 & 228 & 228 & 248 & 238 & 242 & 243\\
 & 248 & 247 & 256 & 246 & 249 & 248\\
 & 248 & 247 & 256 & 246 & 249 & 248\\
 & 391 & 382 & 434 & 417 & 381 & 411\\
 & 418 & 411 & 454 & 436 & 407 & 437\\
 & 418 & 411 & 454 & 436 & 407 & 437\\
 & 473 & 467 & 492 & 472 & 455 & 471\\
 & 473 & 467 & 492 & 472 & 455 & 471\\
 & 479 & 474 & 493 & 473 & 461 & 475\\
 & 479 & 474 & 493 & 473 & 461 & 475\\
 & 486 & 482 & 497 & 477 & 470 & 475\\
\bottomrule
\end{tabular}
\end{table}

\subsubsection{Empirical corrections.~~}
\label{sec:orgheadline19}
Empirical scale factors were determined by fitting the frequencies to the experimental spectrum for S\(_8\).
Note that frequencies are linearly proportional to their corresponding zero-point energies \(E_\textrm{ZPE} = \tfrac{1}{2}h \nu\) and hence this may also be seen as fitting to zero-point energy on a per-mode basis.
The factors were calculated for each functional (Table~\ref{tbl:scale_factors});
scaling the frequencies from PBE0 by 96\% was found to give the best overall fit, and is employed here as the reference "empirically-corrected" method. 
The resulting set of frequencies is illustrated in Fig. \ref{fig:empirical_freqs} alongside the uncorrected and experimental values.
Using this scale factor also gives good agreement (< 4~cm\(^{\text{-1}}\) error) with the stretching frequency of S\(_2\), which was not used in the fit. (Table~\ref{tbl:frequencies})
Least-squares fitting was carried out with the Levenberg-Marquardt algorithm as implemented in Scipy.\cite{Marquardt1963,Scipy2001}

\begin{table}[htb]
\small
\caption{\label{tbl:scale_factors}
Optimal scale factors for exchange-correlation functionals, fitting to ground-state frequencies of S\(_{\text{8}}\)\cite{Chase1998}. Standard deviations \(s\) for the least-squares fit are given over the set of frequencies in units of frequency and their corresponding zero-point energies per sulfur atom.}
\centering
\begin{tabular}{lrrr}
\toprule
Functional & scale factor & \(s\) / cm\(^{\text{-1}}\) & \(s\) / eV (ZPE)\\
\midrule
LDA & 1.0085 & 11.57 & 0.00072\\
PBEsol & 1.0201 & 12.39 & 0.00077\\
PBE0 & 0.9596 & 6.41 & 0.00040\\
B3LYP & 1.0332 & 11.05 & 0.00068\\
\bottomrule
\end{tabular}
\end{table}

\subsection{Equilibrium model}
\label{sec:orgheadline21}
\begin{figure*}
\centering
\includegraphics[width=0.6\textwidth]{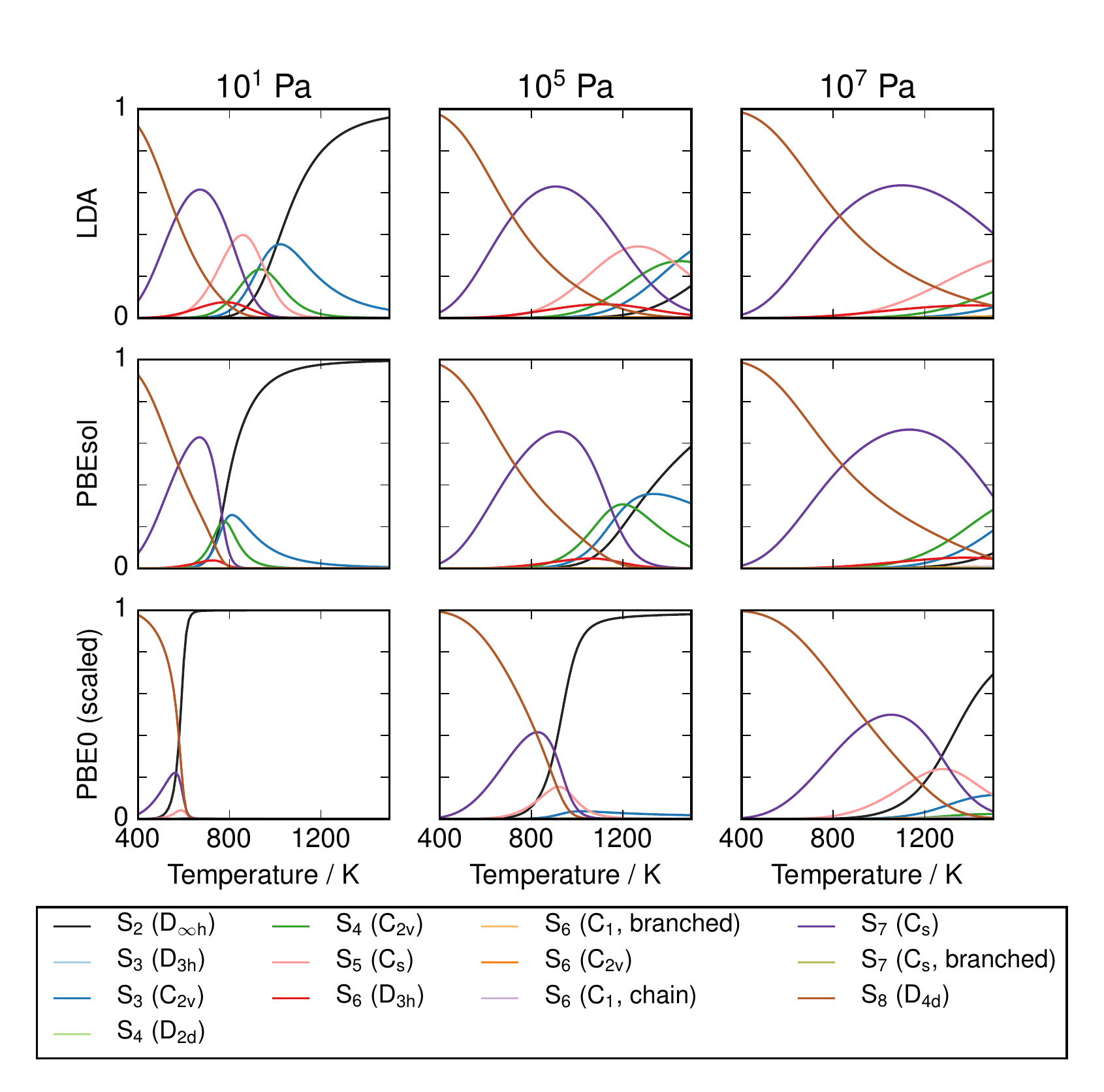}
\caption{\label{fig:composition}
Compositions of modelled S\(_{x}\) mixtures over range of equilibrium temperatures and pressures. 
Results are presented for density functional theory with one local (LDA), one semi-local (PBEsol) and one non-local exchange-correlation functional with empirical corrections. Composition is given in units of atom fraction. It is expected that the most accurate results are obtained using PBE0 with scaled frequencies.}
\end{figure*}

Equilibrium compositions and free energies were computed as a function of temperature and pressure for all the data sets computed (Fig.~\ref{fig:composition}).
There is significant disagreement between the predictions of the local exchange-correlation functionals LDA and PBEsol and the predicted composition from the hybrid functional PBE0, both before and after frequency scaling.
While the "lower-level" calculations predict a diverse mixture of phases, hybrid DFT strongly supports the dominance of S\(_8\) and S\(_2\), at low and high temperatures respectively.
In all cases, this simplicity is strongest at low total pressure.
The other phases which are present in any significant quantity are the cyclic allotropes where \(N\) = 4-7, in the range 600-1000 K.

\begin{figure*}
\centering
\includegraphics[width=\textwidth]{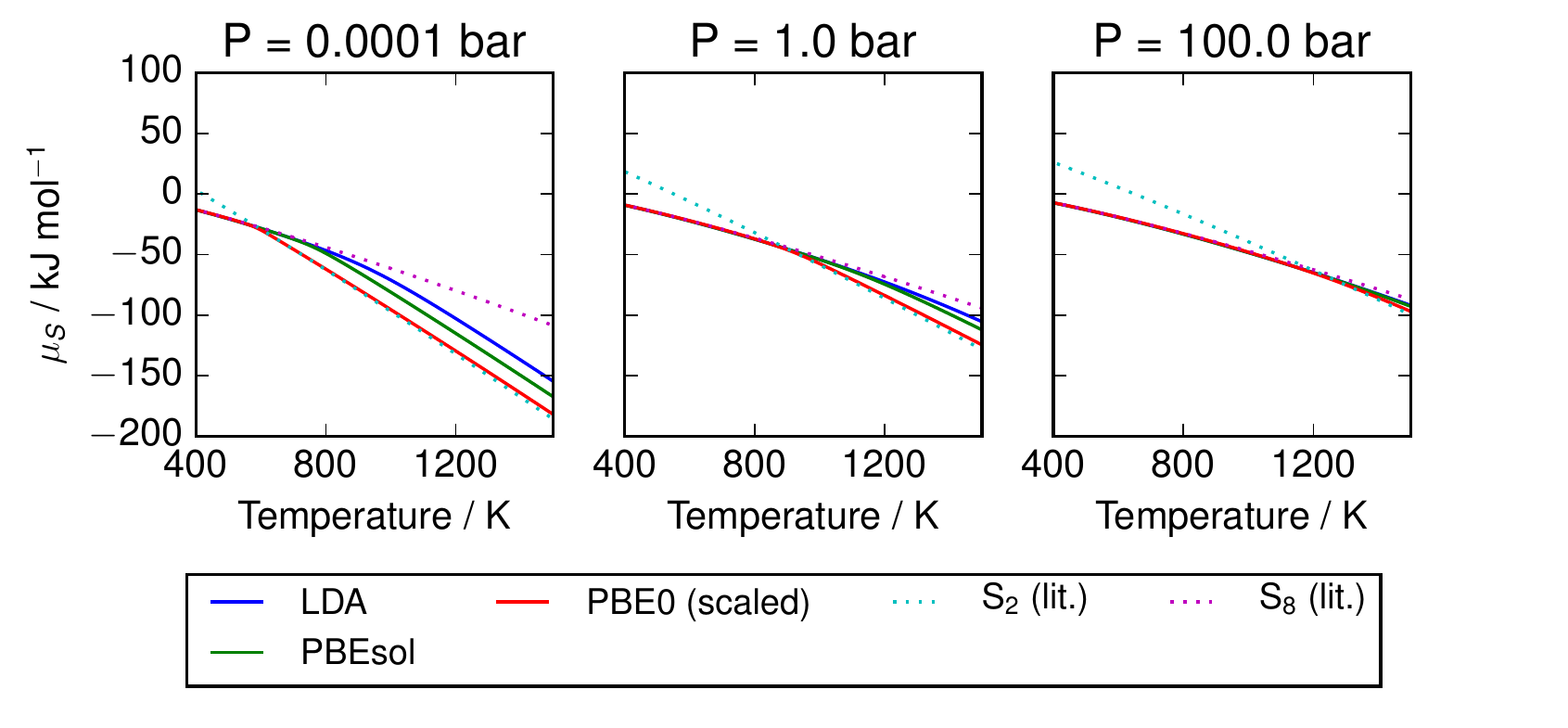}
\caption{\label{fig:mu_functionals}
Chemical potential of S vapours per mole of atoms, given at several pressures according to range of calculation methods. Data for S\(_2\) and S\(_8\) are also provided from the thermochemical literature.\cite{Chase1998} At low pressures, the free energy diverges by more than 50 kJ mol\(^{-1}\) S atoms between the S\(_2\) and S\(_8\) allotropes at high temperatures, while at high pressures there is less variation. Results from hybrid DFT calculations with scaled frequencies closely track the minimal value from the literature, while the local and semi-local exchange correlation functionals diverge from this data due to over-estimation of the formation energy of S\(_2\).}
\end{figure*}

\begin{figure*}
\centering
\includegraphics[width=\textwidth]{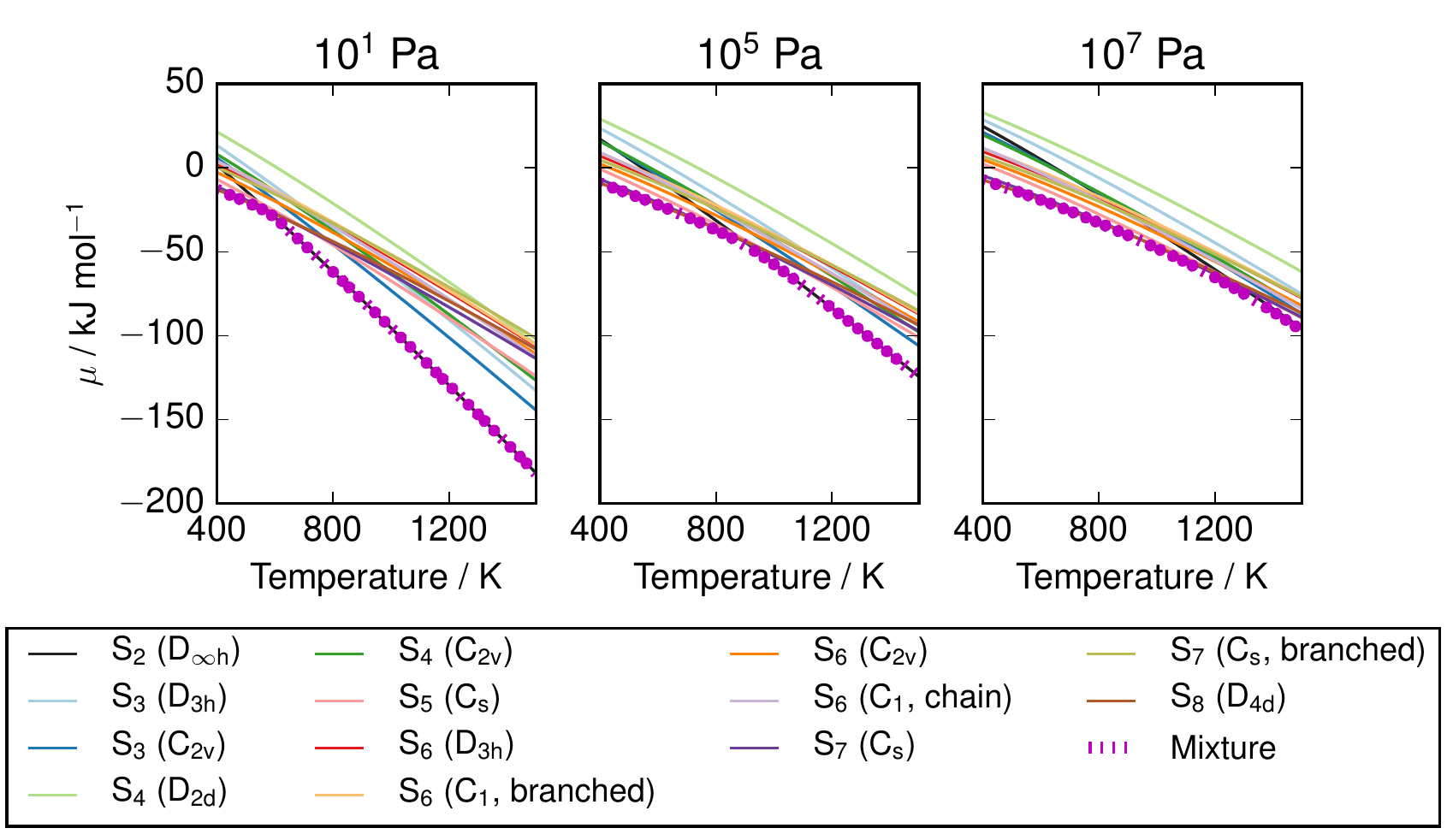}
\caption{\label{fig:mu_contributions}
Chemical potential of S vapours over range of T, P, compared with individual allotropes. The equilibrium mixture is lower in energy than any single allotrope, but in most T/P regimes lies close to the chemical potential of S\(_2\) or S\(_8\). Data from vibrational calculations with PBE0 and empirically-corrected frequencies.}
\end{figure*}

The corresponding free energies are also plotted in Figure~\ref{fig:mu_functionals}; we note that agreement between the methods is much stronger at low temperatures where the mixture is dominated by larger molecules.
This may be an artefact of aligning the free energies of the S\(_{8}\) atoms; divergence in the energies of the smaller molecules leads to the disagreement at high temperatures.
The other trend of note is the presence of a sharp bend in the \(\mu-T\) curve, particularly at low pressure, corresponding to the presence of S\(_2\) molecules.
The point of onset depends on the data source, but the curve for PBE0 with empirical corrections closely tracks the minimum of the two curves from reference data.
This represents a challenge to the formation of a simple parameterised model function, as it suggests the presence of a spike in the second derivative.
Popular parameterisations of thermochemical properties, such as those in the NIST "WebBook", employ multiple temperature regions. 
This is usually viewed as a limitation, as it introduces non-physical discontinuities; with care, they could be aligned to an apparently physical discontinuity in the function.
Taking the PBE0 results with empirical corrections as our preferred model, the free energy of the mixture is plotted with the chemical potentials of its component species on an atomic basis (Fig.~\ref{fig:mu_contributions}).

\begin{table*}[htb]
\small
\caption{\label{tbl:mu_pbe0_scaled}Gibbs free energy of S vapours, tabulated from calculations with PBE0 and empirical corrections, with reference state (H=0) $\alpha$-sulfur at 298.15K. Energies in kJ mol$^{\text{-1}}$, column headers in log$_{\text{10}}$(pressure/Pa). Tables are provided with more values and greater decimal precision in the supplementary information.}
\centering
\begin{tabular}{rrrrrrrrrrr}
\toprule
\multirow{2}{*}{T/K} &  \multicolumn{10}{c}{$\log_{10}(p/\mathrm{Pa})$} \\
 & 1.00 & 1.67 & 2.33 & 3.00 & 3.67 & 4.33 & 5.00 & 5.67 & 6.33 & 7.00\\
\cmidrule(lr){1-1} \cmidrule(lr){2-11}
100 & 4.73 & 4.88 & 5.04 & 5.20 & 5.36 & 5.52 & 5.68 & 5.84 & 6.00 & 6.16\\
150 & 2.29 & 2.53 & 2.77 & 3.01 & 3.25 & 3.49 & 3.72 & 3.96 & 4.20 & 4.44\\
200 & -0.39 & -0.07 & 0.25 & 0.57 & 0.89 & 1.21 & 1.53 & 1.85 & 2.17 & 2.49\\
250 & -3.27 & -2.87 & -2.47 & -2.08 & -1.68 & -1.28 & -0.88 & -0.48 & -0.08 & 0.32\\
300 & -6.34 & -5.86 & -5.39 & -4.91 & -4.43 & -3.95 & -3.47 & -2.99 & -2.51 & -2.03\\
350 & -9.58 & -9.02 & -8.46 & -7.90 & -7.34 & -6.78 & -6.23 & -5.67 & -5.11 & -4.55\\
400 & -12.97 & -12.33 & -11.69 & -11.05 & -10.41 & -9.77 & -9.13 & -8.49 & -7.85 & -7.21\\
450 & -16.50 & -15.77 & -15.05 & -14.33 & -13.61 & -12.89 & -12.17 & -11.45 & -10.73 & -10.01\\
500 & -20.20 & -19.37 & -18.56 & -17.75 & -16.94 & -16.14 & -15.33 & -14.53 & -13.73 & -12.93\\
550 & -24.24 & -23.17 & -22.22 & -21.31 & -20.40 & -19.51 & -18.62 & -17.73 & -16.85 & -15.96\\
600 & -29.74 & -27.46 & -26.12 & -25.03 & -24.01 & -23.01 & -22.03 & -21.05 & -20.08 & -19.11\\
650 & -37.54 & -33.52 & -30.62 & -29.01 & -27.78 & -26.65 & -25.56 & -24.49 & -23.42 & -22.36\\
700 & -45.63 & -41.17 & -36.83 & -33.61 & -31.81 & -30.45 & -29.22 & -28.04 & -26.87 & -25.72\\
750 & -53.78 & -49.00 & -44.23 & -39.63 & -36.36 & -34.48 & -33.03 & -31.71 & -30.43 & -29.18\\
800 & -61.99 & -56.89 & -51.79 & -46.72 & -41.99 & -38.90 & -37.03 & -35.51 & -34.10 & -32.74\\
850 & -70.27 & -64.84 & -59.43 & -54.02 & -48.67 & -44.06 & -41.31 & -39.46 & -37.88 & -36.39\\
900 & -78.59 & -72.85 & -67.11 & -61.38 & -55.67 & -50.16 & -46.04 & -43.61 & -41.79 & -40.15\\
950 & -86.97 & -80.91 & -74.85 & -68.80 & -62.75 & -56.78 & -51.43 & -48.04 & -45.84 & -44.01\\
1000 & -95.39 & -89.01 & -82.64 & -76.26 & -69.90 & -63.57 & -57.48 & -52.84 & -50.06 & -47.98\\
1050 & -103.86 & -97.17 & -90.47 & -83.77 & -77.09 & -70.43 & -63.88 & -58.14 & -54.50 & -52.07\\
1100 & -112.38 & -105.36 & -98.34 & -91.33 & -84.32 & -77.34 & -70.42 & -63.91 & -59.21 & -56.29\\
1150 & -120.94 & -113.60 & -106.26 & -98.93 & -91.60 & -84.29 & -77.03 & -70.00 & -64.26 & -60.68\\
1200 & -129.53 & -121.88 & -114.22 & -106.57 & -98.92 & -91.29 & -83.70 & -76.25 & -69.65 & -65.25\\
1250 & -138.17 & -130.19 & -122.22 & -114.24 & -106.28 & -98.33 & -90.41 & -82.60 & -75.33 & -70.03\\
1300 & -146.84 & -138.54 & -130.25 & -121.96 & -113.67 & -105.40 & -97.16 & -89.01 & -81.23 & -75.04\\
1350 & -155.55 & -146.93 & -138.32 & -129.71 & -121.10 & -112.51 & -103.95 & -95.46 & -87.25 & -80.27\\
1400 & -164.29 & -155.36 & -146.42 & -137.49 & -128.57 & -119.66 & -110.77 & -101.95 & -93.36 & -85.72\\
1450 & -173.06 & -163.81 & -154.56 & -145.31 & -136.07 & -126.84 & -117.63 & -108.49 & -99.53 & -91.33\\
\bottomrule
\end{tabular}
\end{table*}

\begin{figure}[htb]
\centering
\includegraphics[width=8.3cm]{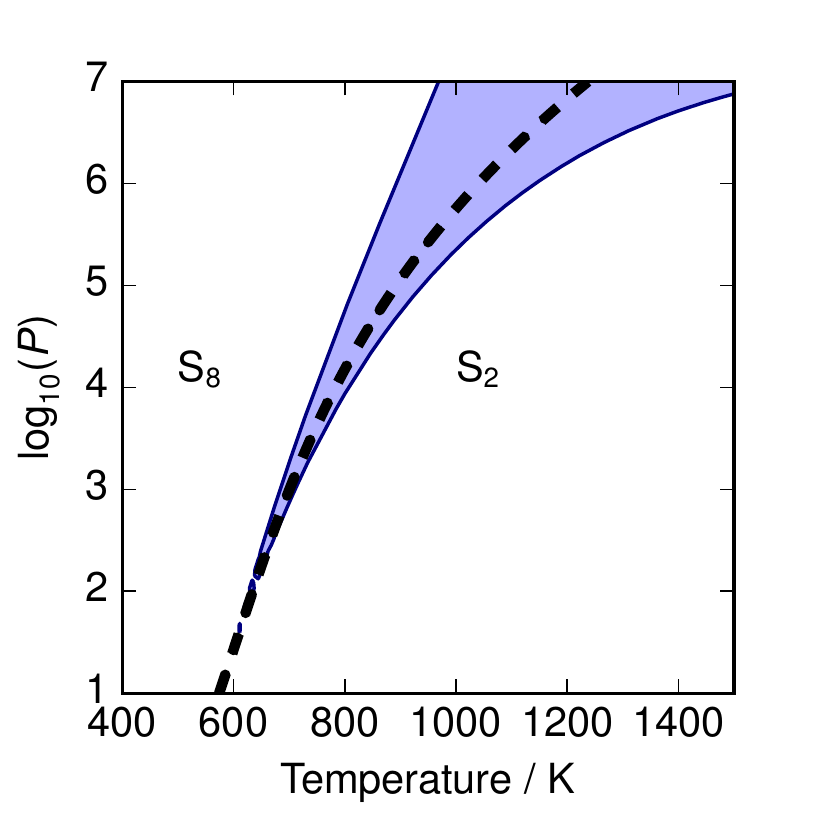}
\caption{\label{fig:surface}
Temperature-pressure map of approximations to free energy of mixture. At dashed line \(\tfrac{1}{2} \mu_{\text{S}_2} = \tfrac{1}{8} \mu_{\text{S}_8}\); in shaded region the error in chemical potential \(\mu\) associated with assuming a single phase S\(_2\) or \(S_8\) exceeds 1 kJ/mol S atoms; in unshaded regions the corresponding single-phase free energy is close to the energy of the mixture.}
\end{figure}

\begin{figure}[htb]
\centering
\includegraphics[width=8.3cm]{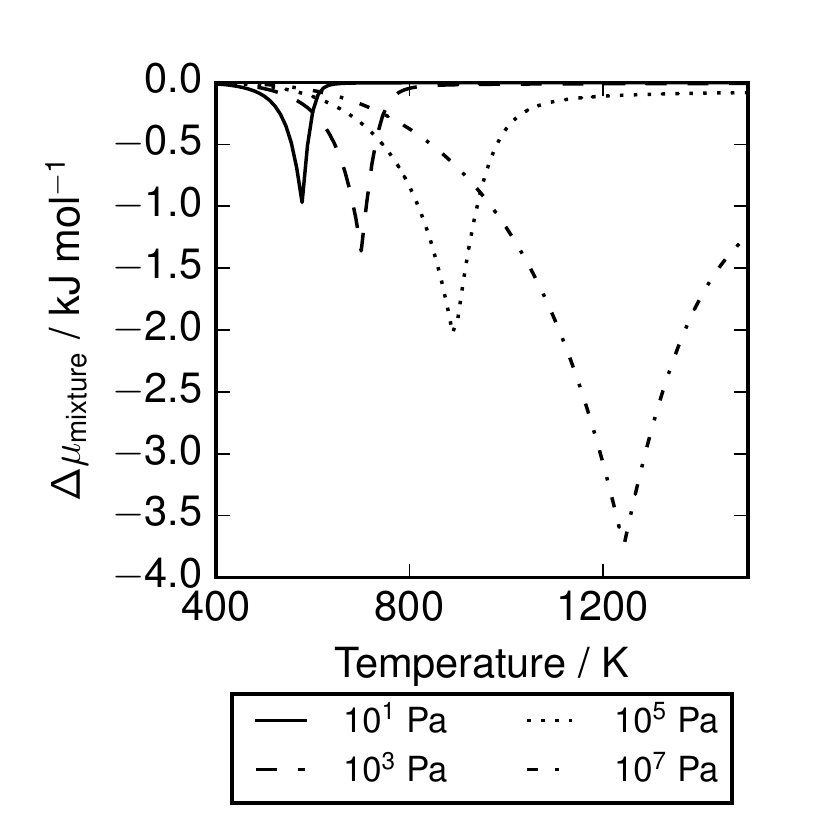}
\caption{\label{fig:mu_mix_contribution}
Depression in chemical potential of sulfur vapour \(\mu_{\mathrm{S}}\) due to mixing and presence of minor allotropes. \(\Delta \mu_{\mathrm{mixture}} = \mu_{\mathrm{S}} - \mathrm{min}\left( \frac{\mu_{\mathrm{S}_2}}{2}, \frac{\mu_{\mathrm{S}_8}}{8}\right)\)}
\end{figure}

The depression in free energy due to mixing of allotropes and presence of minor components can be quantified by subtracting the chemical potential of the mixture from the minimum of the chemical potentials of the majority components S\(_2\) and S\(_8\).
The resulting plot (Fig.~\ref{fig:mu_mix_contribution}) shows that this has an impact ranging from around 1--4 kJ mol\(^{-1}\), depending on the pressure.
This is illustrated as a contour plot in Fig. \ref{fig:surface}; within each unshaded region a single-phase model is adequate to within 1 kJ mol\(^{-1}\) S atoms.

\subsection{Parameterisation}
\label{sec:orgheadline22}

For convenience, a parameterised fit has been generated for the chemical potential of S over the T, P range 400--1500K, 
10\(^{\text{0}}\)--10\(^{\text{7}}\) Pa, incorporating an error function "switch" between S\(_{\text{2}}\) and S\(_{\text{8}}\) dominated regions and a Gaussian correction for the free energy depression where there is substantial mixing of phases. In eV per S atom, for $T$ in K, the form of the parameterisation is
\begin{align}
\mu_{\mathrm{S}}(T,P) &= 
\frac{1}{2} \left[ \mathrm{erfc}\left( \frac{T - T_{tr}}{w} \right) \frac{\mu_{\mathrm{S}_8}}{8} + 
\left( \mathrm{erf} \left(\frac{T - T_{tr}}{w}\right) + 1 \right) \frac{ \mu_{\mathrm{S}_2}}{2} \right] \nonumber \\
&\hphantom{=}  - a(P) \exp \left( - \frac{\left(T - T_{tr} + b \right)^2}{2 c^2} \right)
\end{align}
where \(\mu_{\mathrm{S}_8} (T,P) = \num{7.620e-1} - \num{2.457e-3} T - \num{4.012e-6} T^2 + \num{1.808e-9} T^3 - \num{3.810e-13} T^4 + k_B \ln\left(\frac{P}{\mathrm{1 bar}}\right)\), 
\(\mu_{\mathrm{S}_2} (T,P) = \num{1.207} - \num{1.848e-3} T - \num{8.566e-7} T^2 + \num{4.001e-10} T^3 - \num{8.654e-14} T^4 + k_B \ln\left(\frac{P}{\mathrm{1 bar}}\right)\).
\(T_{tr}\), the transition temperature obtained by solving \(\tfrac{1}{2} \mu_{\mathrm{S}_2} = \tfrac{1}{8} \mu_{\mathrm{S}_8}\) is approximated by the polynomial \(T_{tr} = \num{5.077e2} + \num{7.272e1}\log_{10}P  - \num{8.295e0}(\log_{10}P)^2 + \num{1.828e0}(\log_{10}P)^3\). 
The height of the Gaussian correction \(a(P) = \num{1.414e3} - \num{2.041e2}\log_{10}P + \num{6.663e1}(\log_{10}P)^2\), and the more arbitrarily assigned width and offset parameters \(b=10\), \(c = 80\), \(w = 100\).

\begin{figure}[htb]
\centering
\includegraphics[width=8.3cm]{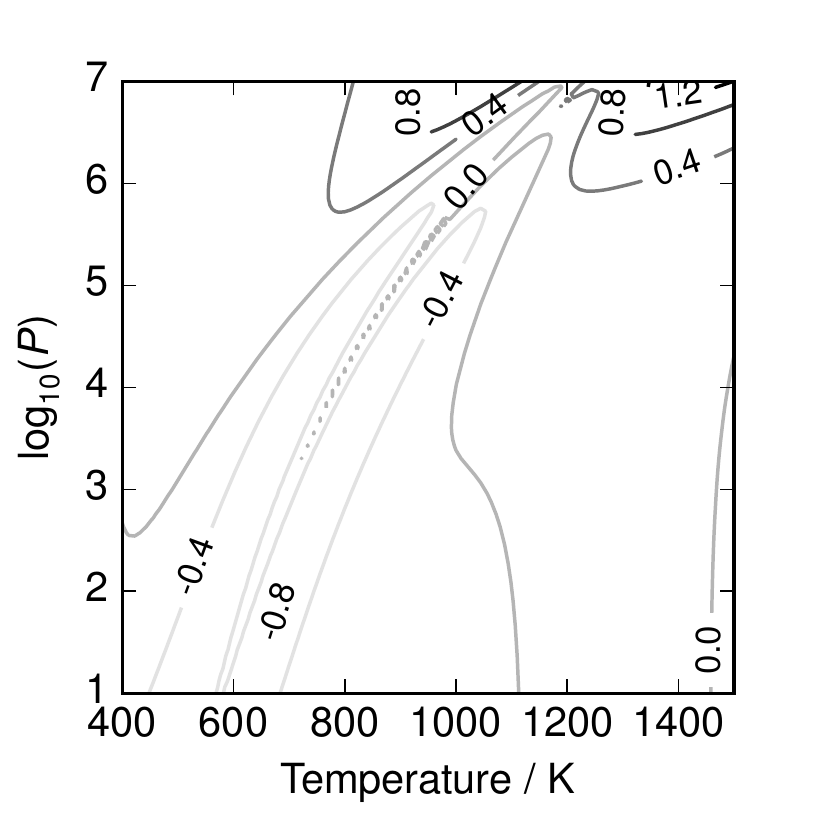}
\caption{\label{fig:param_error}
Error of parameterisation in kJ mol\(^{-1}\). Error is reduced to less than 1 kJ mol\(^{-1}\), but is highly non-uniform. Parameterisation is recommended for convenient application over wide T--P ranges; the full equilibrium solution is required to correctly capture fine detail.}
\end{figure}

It is noted that this parameterisation contains many fitting parameters; however, given its physically-motivated form the resulting function is smooth and well-behaved over the region studied, while the fits to \(\mu_{\mathrm{S}_2}\), \(\mu_{\mathrm{S}_8}\) and \(T_{tr}\) have some value in their own right. The fitting error is plotted in Fig.~\ref{fig:param_error}, and while somewhat irregular remains below 1 kJ mol\(^{-1}\).

\section{Conclusions}
\label{sec:orgheadline24}
The chemical potential of sulfur vapours has been studied by solving the thermodynamic equilibrium of 13 gas-phase allotropes, including the dominant components S\(_2\) and S\(_8\). 
Thermochemical data was obtained from first-principles calculations and corrected with an empirical scaling factor for the vibrational frequencies. 
The transition between these dominating phases is highly pressure-dependent, and the free energy is further depressed at the transition temperature by the presence of additional phases, especially at elevated pressures. 
Selection of an inappropriate gas phase can lead to errors of the order 50 kJ mol\(^{-1}\) atoms, while the minor phases contribute free energy of the order 1 kJ mol\(^{-1}\) atoms.
The resulting chemical potential data is made available through tabulated data, a parameterised model with error of the order 0.5 kJ mol\(^{-1}\) atoms and through open-source code; the reference energy is compatible with the NIST-Janaf thermochemical tables for the solid \(\alpha\)-sulfur phase.\cite{Chase1998}
This phase is frequently used as a reference state for thermodynamic studies of defects and stability in metal chalcogenides;
the application of this gas-phase potential may allow such studies to examine a wide range of reactions involving sulfur vapours, taking into account the equilibrium within the vapour phase.
The selection of appropriate chemical potentials is also critical for the development and interpretation of phase diagrams.

\section{Data Access Statement}
\label{sec:orgheadline25}
The reference implementation of this model, complete with Python 2.7 code to generate all the plots in this paper as well as tabulated data in the form of Table~\ref{tbl:mu_pbe0_scaled}, is available online at \url{https://github.com/WMD-Bath/sulfur-model} and a snapshot of the code at the point of submission of this article is hosted by Zenodo and available with the DOI: \texttt{10.5281/zenodo.28536}.
In addition, full tables are provided with this paper in the ESI\(^{\dag}\) for the composition, enthalpy and chemical potential from the calculations with PBE0 and empirical corrections; 
one set of enthalpy and chemical potential data follows Table~\ref{tbl:mu_pbe0_scaled} and uses the enthalpy of \(\alpha\)-S as a reference energy (for use with other tabulated data) while the other employs the ground state of S\(_8\) as a reference energy (for use with first-principles calculations.)
The code and its dependencies are Free Software, using a range of licenses.
Input and output files from DFT calculations with FHI-aims have been deposited with Figshare and are available with the DOI: \texttt{10.6084/m9.figshare.1513736}.
A set of data generated during the evolutionary search, consisting of candidate structures and the DFT energies used to rank them,
 has been deposited with Figshare and is available with the DOI: \texttt{10.6084/m9.figshare.1513833}.

\section{Acknowledgements}
\label{sec:orgheadline26}

The authors thank J. M. Skelton and J. M. Frost for useful discussions.
USPEX/VASP calculations with PBE were carried out using the University of Bath's High Performance Computing facilities.
Hybrid DFT calculations were carried out using ARCHER, the UK's national high-performence computing service, via our membership of the UK's HPC Materials Chemistry Consortium, which is funded by EPSRC (grant no. EP/L000202).
A.J.J. is part of the EPSRC Doctoral Training Center in Sustainable Chemical Technologies (grant no. EP/G03768X/1).
The contribution of D.T. was supported by ERC Starting Grant 277757.

\bibliography{sulfur}
\bibliographystyle{rsc} 

\appendix

\section{Proof that \(G = \lambda\)}
\label{sec:orgheadline27}
\label{SEC:lagrange_gibbs_proof}
We define the molar Gibbs free energy of sulfur atoms in a molecular gas mixture as
\begin{align}
\hat{G}_\text{S} (T,P) &= \frac{\sum\limits_{i=1}^N n_i \mu_i}{b} = \sum\limits_{i=1}^N \frac{n_i}{b} \mu_i \\
\intertext{and substitute in (\ref{eq:phi_frac})}
\hat{G}_\text{S} (T,P) &= \sum^N_{i=1} \left[ \frac{\mu_i \Phi_i}{\sum^N_{j=1} a_j\Phi_j} \right]. \\
\intertext{
(Notation the same as Sec. \ref{SEC:eqm_derivation}). From (\ref{eq:eqm-cond1}), $\mu_i = a_i \lambda$ and hence}
\hat{G}_\text{S} (T,P) &= \frac{\sum^N_{i=1}a_i \lambda \Phi_i}{\left( \sum^N_{j=1} a_j \Phi_j \right)} = \lambda
\end{align}

\end{document}